\documentclass[aps, onecolumn, groupedaddress,nofootinbib,longbibliography, notitlepage]{revtex4-1}

\usepackage{graphicx}
\usepackage{xcolor}
\usepackage[separate-uncertainty=true, multi-part-units=single]{siunitx}

\begin{document}


\title{Pair dispersion in inhomogeneous turbulent thermal convection}


\author{Olivier Liot}
 \email[]{olivier.liot@imft.fr}
 \altaffiliation[Presently at ]{Institut de M\'ecanique des Fluides de Toulouse, Toulouse, France}
 \author{David Martin-Calle}
 \altaffiliation[Presently at ]{Institut Lumi\`ere Mati\`ere, CNRS, 8 rue Ada Byron, Villeurbanne, France}
\author{Am\'elie Gay}%
\altaffiliation[Presently at ]{IRPHE - UMR 7342, Technop\^ole de Ch\^ateau-Gombert, 49, rue Joliot Curie - B.P. 146, 13384 Marseille Cedex 13, France}
\author{Julien Salort}
\author{Francesca Chill\`a}
\author{Micka\"el Bourgoin}
\affiliation{Univ Lyon, ENS de Lyon, Univ Claude Bernard, CNRS, Laboratoire de Physique, F-69342 Lyon Cedex 7, France}
%


\date{\today}

\begin{abstract}
{Due to large scale flow inhomogeneities and the effects of temperature, turbulence small-scale structure in thermal convection is still an active field of investigation, especially considering sophisticated Lagrangian statistics. 
Here we experimentally study Lagrangian pair dispersion (one of the canonical problems of Lagrangian turbulence) in a Rayleigh-B\'enard convection cell. 
A sufficiently high temperature difference is imposed on a horizontal layer of fluid to observe a turbulent flow. 
We perform Lagrangian tracking of sub-millimetric particles on a large measurement volume including part of the Large Scale Circulation (LSC) revealing some large inhomogeneities. 
Our study brings to light several new insights regarding our understanding of turbulent thermal convection: 
(i) by decomposing particle Lagrangian dynamics into the LSC contribution and the turbulent fluctuations, we highlight the relative impact of both contributions on pair dispersion; 
(ii) using the same decomposition, we estimate the Eulerian second-order velocity structure functions from pair statistics and show that after removing the LSC contribution, the remaining statistics recover usual homogeneous and isotropic behaviours which are governed by a local energy dissipation rate to be distinguished from the global dissipation rate classically used to characterise turbulence in thermal convection; 
and (iii) we revisit the super-diffusive Richardson-Obukhov regime of particle dispersion and propose a refined estimate of the Richardson constant.}
\end{abstract}


\maketitle


\section{Introduction} 

Various natural and industrial flows are buoyancy-driven. 
Atmospheric or oceanic dynamics, Earth's mantle flows and heat exchangers are some examples where the natural thermal convection -- without mechanical forcing -- has a dramatic influence on the flow. 
The main model system is the Rayleigh-B\'enard cell. 
A horizontal layer of fluid, confined in a cell is cooled from above and heated from below. 
If the thermal forcing is large enough, the flow is turbulent and the fluid is well mixed. 
A consequence is that the temperature is nearly homogeneous on average, and thermal gradients are confined in boundary layers, close to the horizontal plates. 
Boundary layers are thermally unstable, which leads to the emission of coherent structures, named thermal plumes. 
Despite the significant progress of the modelling of turbulent thermal convection in the past decade \cite{grossmann2000,ahlers2009,lohse2010}, the link between these structures and the global heat flux is still not fully understood. 
Examples of open questions are the dynamics of the turbulent plumes \cite{chilla2012}, the interactions between the mean flow and the boundary layers, and the influence of the inhomogeneity on the heat transport.

The system is controlled by two non dimensional parameters. 
(i) The Rayleigh number, $\mathrm{Ra}$, is the balance of the buoyancy effects and the dissipative ones. 
It accounts for the thermal forcing,
\begin{equation}
\mathrm{Ra}=\frac{g\alpha\Delta T H^3}{\nu\kappa},
\end{equation}
where $g$ is the acceleration due to gravity, $\Delta T$ is the temperature difference between the cooling and the heating plates, $H$ is their separation distance, $\alpha$ represents the thermal expansion coefficient of the fluid, $\nu$ the fluid's kinematic viscosity and $\kappa$ its thermal diffusivity. 
(ii) The Prandtl number, $\mathrm{Pr}$, compares the two dissipative processes (thermal and viscous diffusion): 
\begin{equation}
\mathrm{Pr}=\frac{\nu}{\kappa}.
\end{equation}
The response of the system is estimated by the Nusselt number, $\mathrm{Nu}$, which compares the heat flux through the convection cell to the purely diffusive one,
\begin{equation}
\mathrm{Nu}=\frac{Q H}{\lambda\Delta T},
\end{equation}
where $Q$ is the heat flux and $\lambda$ the thermal conductivity of the fluid.

High-resolution spatio-temporal Lagrangian measurements in turbulent flows are now possible thanks to improved digital imaging techniques and computing resources \cite{la_porta2001,voth2002}. 
Flows with important mixing and transport properties deserve to be studied with a Lagrangian approach \cite{fox2003}.
Moreover, the Lagrangian point of view is relevant to stochastic models which describe some aspects of the turbulence, such as finite-Reynolds effects \cite{sawford1991,pope2002}, and intermittency \cite{arneodo2008}. 
Important transport properties of turbulent flows are related to multi-particle dispersion \cite{falkovich2001}, among which pair dispersion is the most fundamental and has been pioneered by Sir Richardson \cite{richardson1926}. 
The way two particles go away from each other has been largely studied in turbulent flows simultaneously experimentally, numerically and theoretically (see e.g. the review articles by Sawford \cite{sawford2001} and Salazar \emph{et al.} \cite{salazar2009}). 
It has important applications including oceanic plankton or atmospheric pollutant dispersion \cite{bourgoin2006,bourgoin2015} for which thermal convection is a key ingredient. 
Three regimes can be distinguished. 
(i) For time scale shorter than $t^*=(\Delta_0^2/\epsilon)^{2/3}$, where $\Delta_0$ is the initial separation, as represented in Fig.~\ref{dispersion_figure}, and $\epsilon$ the kinetic energy dissipation rate, a ballistic regime is observed. 
It is called the Batchelor regime, and the squared pair dispersion $D^2_{\Delta_0}(t)$ (as defined in Eq.~\ref{eq:pairdispersion}) scales as $t^2$ \cite{batchelor1950}. 
(ii) For $t^*<t<T_L$, where $T_L$ is the Lagrangian correlation time, the super-diffusive or Richardson-Obukhov regime appears and $D^2_{\Delta_0}(t)\propto t^3$ \cite{richardson1926}. 
(iii) For $t>T_L$, a diffusive regime is reached and $D^2_{\Delta_0}(t)\propto t$ \cite{bourgoin2006}.

Nevertheless these three regimes are poorly investigated in turbulent thermal convection, contrary to isothermal turbulence. 
{Indeed, the presence of thermal plumes and the strong inhomogeneities of the flow should lead to turbulent statistics far from the usual Homogeneous and Isotropic Turbulence (HIT) framework.} 
To our knowledge, few experimental works were reported.
The first one used a single large particle with embedded thermistors to combine Lagrangian temperature and velocity measurements \cite{gasteuil2007,liot2016-1,liot2017}. 
The second one \cite{ni2012,ni2013} used sub-millimeter particles, focusing on the very center of convection cells where the flow is quite homogeneous and isotropic. 
Recently we reported on a Lagrangian study of the velocity and acceleration statistics in a large measurement volume of a Rayleigh-B\'enard cell where the mean flow is highly inhomogeneous \cite{liot2016}. 
Very recently, a study focused on the aspect ratio influence on Lagrangian statistics \cite{kim2018}. 
Some experimental studies have been performed in rotating Rayleigh-B\'enard turbulence \cite{rajaei2016a,rajaei2016b,rajaei2017,rajaei2018}. 
Moreover, heat flux and particle dispersion were numerically studied by Schumacher and co-workers all along the decade. 
Among these previous studies, the pair dispersion was explored numerically in thermally-driven flows \cite{schumacher2008,schneide2018}. 
The Batchelor and Richardson-Obukhov regimes were observed. 
In experimental studies, mostly the case of initial separations close to the dissipative length-scale was investigated \cite{ni2013}, with some evidence of the Richardson regime.

{The purpose of this paper is to explore how flow inhomogeneity observed in a wide central zone of a thermal flow affects the turbulence. 
We deliberately place this study out of the HIT scope and address the role of the Large Scale Circulation (LSC) on Lagrangian transport.} 
The Lagrangian tracking of particle pairs in a turbulent Rayleigh-B\'enard cell is presented. 
A significant part of the LSC is observed thanks to the size of the measurement volume. 
The growth of the mean-square separation of particles is analysed. 
The main aspect of this article concerns the influence and the modelling of the inhomogeneous mean flow on pair dispersion. 
We also discuss the estimation of the Richardson constant in the Richardson-Obukhov regime.

\section{Experimental setup and measurement techniques} 
\label{section:setup}

The convection cell consists of an octagonal{-shaped} setup with {eight}
transparent polymethylmetacrylate {vertical} walls. 
The two horizontal plates are made of anodised aluminium. 
They are \SI{40}{cm} in diameter, {tangent to the octagon made by the walls} and vertically spaced \SI{30}{cm} apart. 
A  {custom-made \SI{40}{cm} in diameter} spiral electrical resistance imposes the heat flux from the bottom while the top plate temperature is regulated by a glycol circulation pump. 
{The temperature of the plates is monitored using eight PT100 temperature sensors -- four inside each plate.} 
The working fluid is deionised and degased water, {with a density $\rho_f = \num{0.99}$}. 
The imposed heat power is \SI{800}{W}, the mean temperature is fixed to \SI{40}{\celsius} and the temperature difference reaches $\Delta T=\SI{19.2}{K}$. 
The corresponding control parameter values are $\mathrm{Ra}=2.0\times10^{10}$ and $\mathrm{Pr}=4.3$. 
The Nusselt number is consistent with the Grossmann-Lohse theory \cite{grossmann2000,stevens2013}, and is $\mathrm{Nu}=139$ (after removing 15\,\% of thermal losses). 
In these conditions the flow is turbulent. 
The subsequent Large Scale Circulation (LSC) consists in a convection roll confined between two diametrically opposite sides of the octagonal cell. 
{We observe using shadowgraphs that} the convection roll orientation can change spontaneously {with a typical time of few hours.} 
{Nevertheless, w}e do not observe such reversals on the time scale of our acquisitions {(\SI{180}{s}, which is confirmed by the shape-similar velocity distributions of each run \cite{liot2016}.} 
However, LSC sloshing {in the $x$-direction} is observed during the acquisitions \cite{liot2016}.


\begin{figure}[htp]
\begin{center}
\includegraphics[width=7cm]{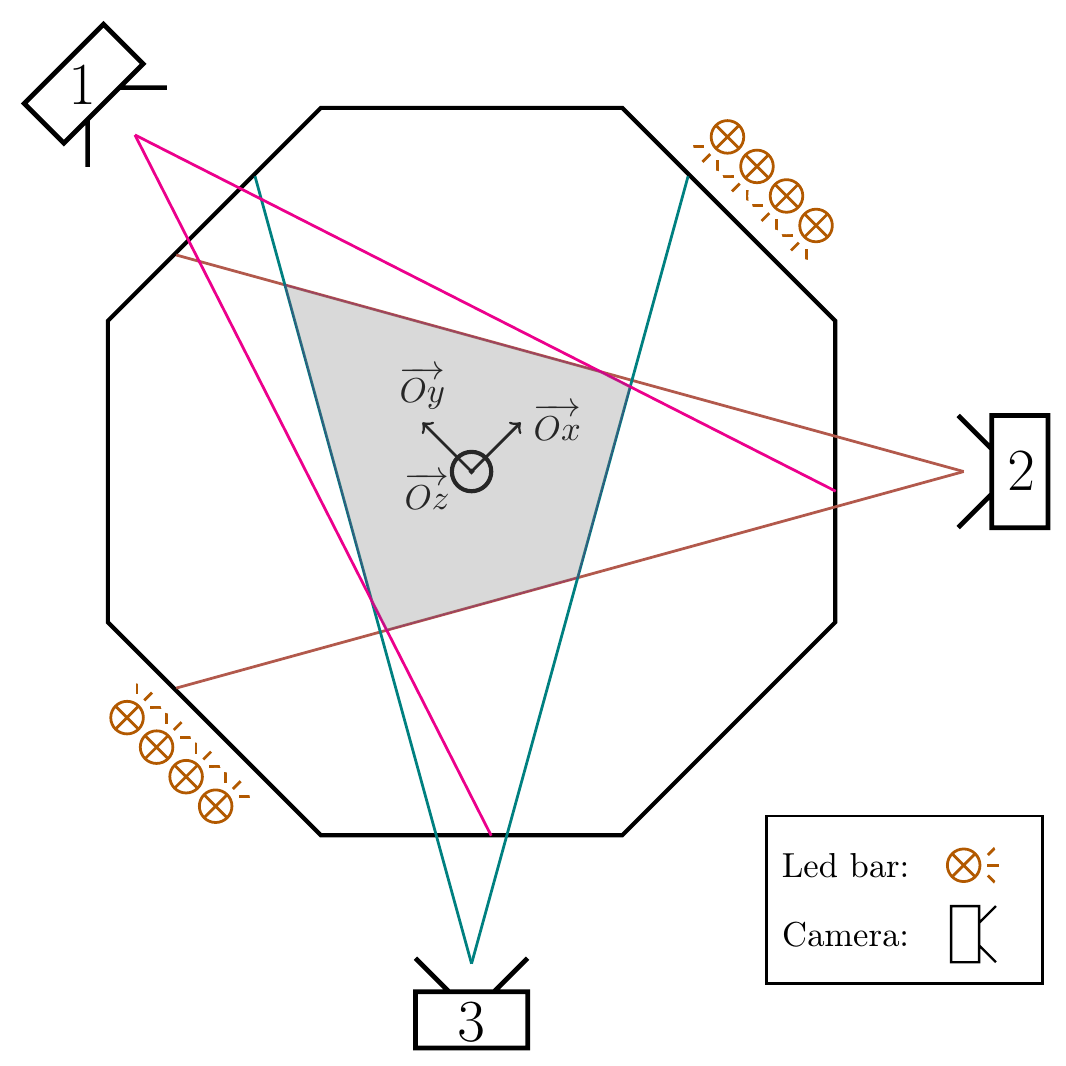}
\caption{\textcolor{black}{Convection cell and camera position viewed from above. 
The gray zone corresponds to the field of view observed by the three cameras. 
More details about dimensions and camera specifications are in the text.
}}
\label{tolosa}
\end{center}
\end{figure}

The flow is seeded with thousands of polystyrene particles. 
Their diameter is $d=\SI{250}{\micro\metre}$ and their density is $\rho_p=\num{1.03}$. 
{The Stokes number in a turbulent flow can be estimated as \cite{qureshi2008,xu2008}:}
\begin{equation}
    \mathrm{St}=\frac{\rho_p}{18\rho_f}\left(\frac{d}{\eta}\right)^2.
\end{equation} 
\noindent {In our case we get $\mathrm{St}\approx 7\times 10^{-3}$, which means that particles can be considered as tracers.
Nevertheless we observe a decrease of suspended particle concentration on a typical time of few dozen minutes, probably due to the entrapment by the thermal boundary layers. 
This problem is solved by splitting the acquisitions into six runs with re-seeding between each one.} 

\begin{figure}[htp]
\begin{center}
\includegraphics[width=7cm]{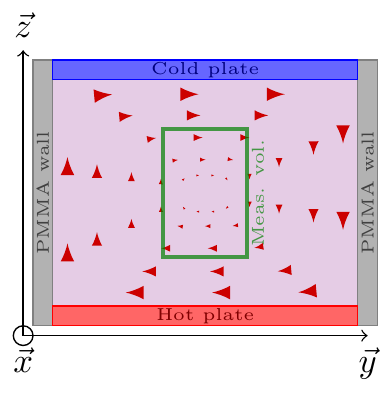}
\caption{Scheme of the convection roll and the measurement volume captured by the experimental acquisition setup, compared to the whole convection cell.}
\label{roll}
\end{center}
\end{figure}

The fluid volume is illuminated by two sets of four vertical LED bars of \SI{864}{lumens}, {put in front of two different vertical walls, as shown in Fig.~\ref{tolosa}.}
{Moreover, to improve the particle visibility, the vertical wall on the far side of the tank from where the camera is situated is covered with black paper. 
Thus only three faces are still available to place the cameras.} 
{To avoid any angle between the camera lens and the wall and to have the largest volume captured by all the cameras, we choose to use only one camera per free face. 
Consequently, t}hree cameras are positioned in the horizontal plane situated at mid-height of the walls, at polar angles $\psi=\SI{0}{\degree}$, \SI{135}{\degree} and \SI{270}{\degree}, as illustrated in Fig.~\ref{tolosa}).
The measurement volume where we perform 3D Lagrangian tracking is about \SI{11}{cm} per side, \SI{17}{cm} high and is centered in the cell (see Fig.~\ref{tolosa} and \ref{roll}). 
The cameras have a resolution of $1088\times 2048$ pixels$^2$. 
The maximum resolution is \SI{6}{pixels} per Kolmogorov length, defined as $\eta = (\nu^3/\langle\epsilon\rangle)^{1/4}$ where $\langle\epsilon\rangle$ is the average mass rate of kinetic energy dissipation, estimated as \cite{shraiman1990}:
\begin{equation}
\langle\epsilon\rangle = \frac{\nu^3}{H^4} \mathrm{Ra}\mathrm{Pr}^{-2}(\mathrm{Nu}-1).
\label{eq:epsilon}
\end{equation}

The acquisitions are split in six independant runs of 180\,s. 
The total measurement time is about 2880 times the Kolmogorov time -- $\tau_\eta = \sqrt{\nu/\langle\epsilon\rangle}\approx0.36$\,s -- and 575 times the free-fall time -- $\tau_{f}=\sqrt{H/g\alpha\Delta T}$.
We can therefore consider that the flow is fairly averaged as long as no LSC orientation change occurs.
The sampling frequency is fixed to \SI{200}{Hz} (about 70 times the dissipative time scale), which ensures a good resolution of small scale Lagrangian dynamics with sufficient oversampling to filter noise on reconnected tracks \cite{voth2002}. 
A Gaussian kernel is used to filter the trajectories and their temporal derivatives; its width ($0.3\,\tau_\eta$) does not affect the smallest turbulent scales.

\begin{figure}[ht]
\begin{center}
\includegraphics[width=15cm]{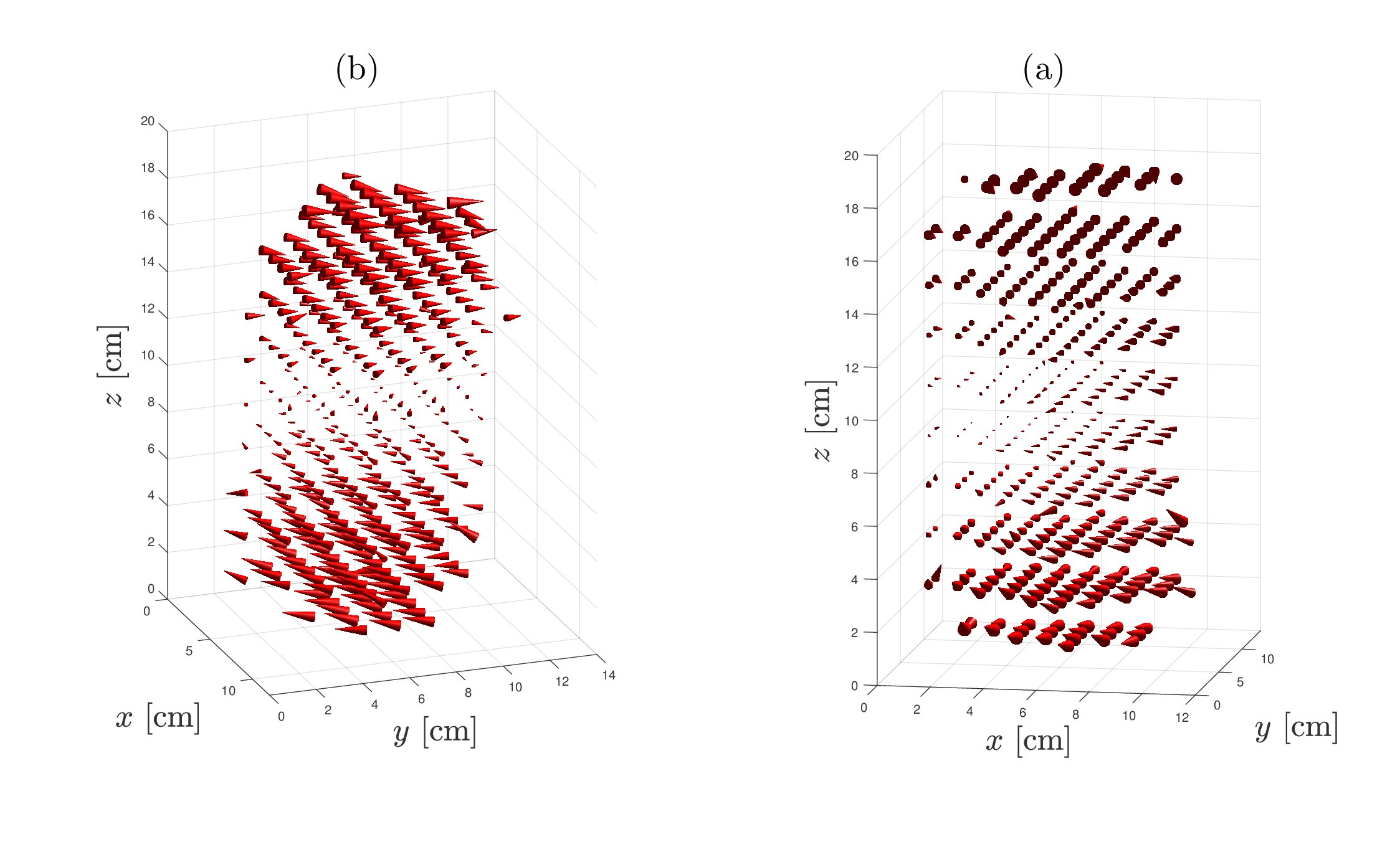}
\caption{Three-dimensional visualisations of the mean velocity field $\protect\overrightarrow{v^E}=v_x^E\protect\overrightarrow{x}+v_y^E\protect\overrightarrow{y}+v_z^E\protect\overrightarrow{z}$. 
The two plots corresponds to two different viewing angles.
The arrow size is proportional to velocity magnitude, going from 0 and \SI{12}{mm/s}.}
\label{vy_3D}
\end{center}
\end{figure}

The measurement volume is wide enough to capture a significant part of the LSC. 
Figure~\ref{roll} presents the part of the convection roll captured by the measurement volume within the convection cell. 
To visualise the inhomogeneity of the flow we compute pseudo-Eulerian maps from the Lagrangian data. 
They are averaged over a spatial grid, and the resulting velocity field is interpolated. 
We call $v^E_k$ ($k=x,y,z$) the resulting average velocity. 
Figures~\ref{vy_3D}~(a) {and (b)} show {three-dimensional representations of the average velocity vectors field, with two distinct viewing angles}. 
{In the $y$-direction,} we observe a very inhomogeneous flow with large positive velocities at the top of the measurement volume, large negative ones at the bottom and null velocities at the centre. 
The other horizontal average velocity component ($v_x^E$) and the vertical one ($v_z^E$) are both nearly null, {as highlighted in Fig.~\ref{vy_3D}~(b) and (a) respectively}. 

\section{Pair dispersion and inhomogeneity} 

\begin{figure}[htp]
\begin{center}
\includegraphics[width=6cm]{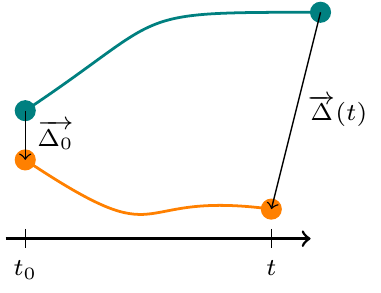}
\caption{Sketch of the pair dispersion principle. At time $t_0$, two particles are separated with an initial distance $\left|\left|\protect\overrightarrow{\Delta_0}\right|\right|$ while at time $t > t_0$, they are separated by $\left|\left|\protect\overrightarrow{\Delta}\right|\right|$.}
\label{dispersion_figure}
\end{center}
\end{figure}

\begin{figure}[ht]
\begin{center}
\includegraphics[width=\textwidth,clip,trim=0 19.7cm 0 0]{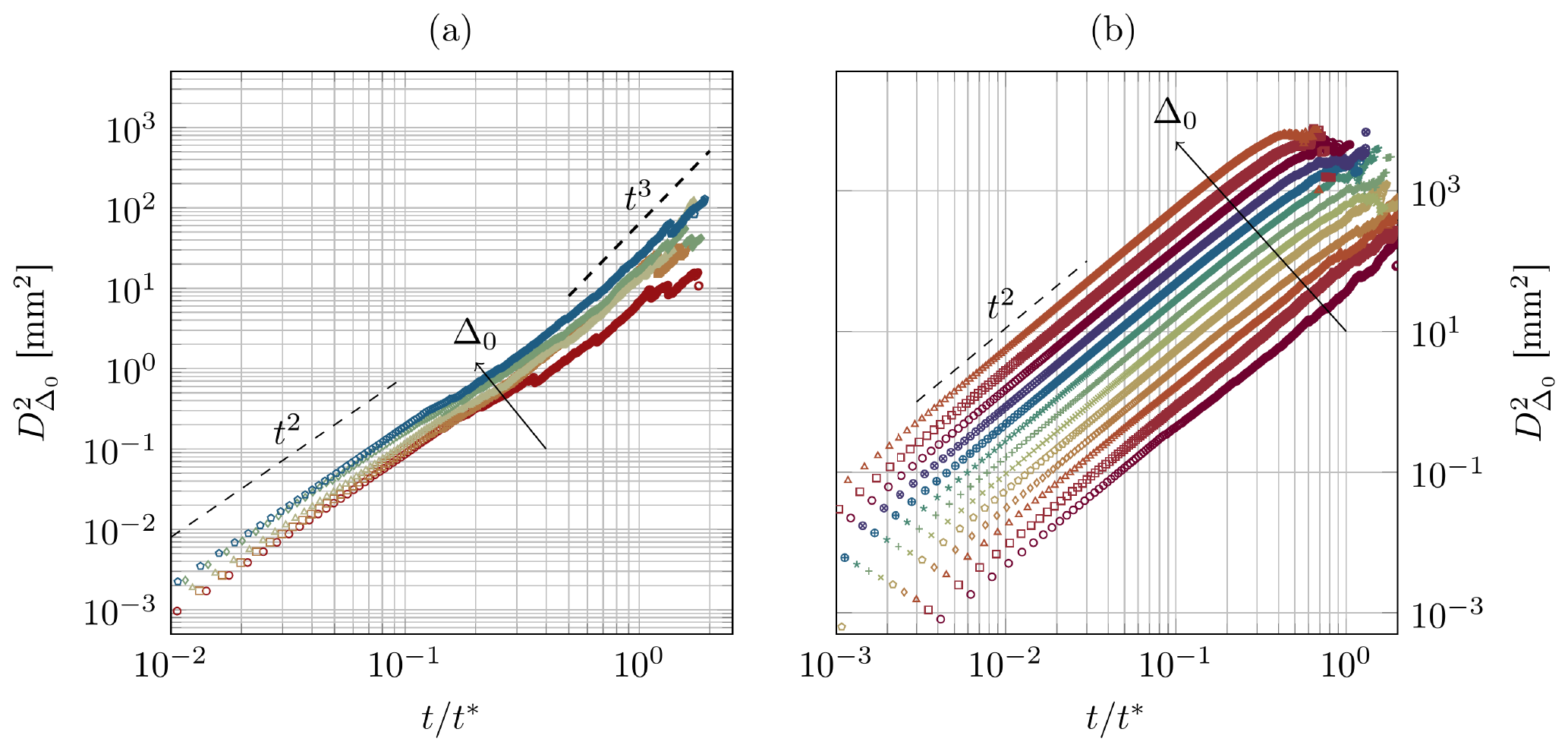}
\caption{Pair dispersion for (a) short and (b) larger initial separations $\Delta_0$ ranging in \{1.9\,mm, 2.1\,mm, 2.3\,mm, 2.6\,mm, 2.9\,mm\} and \{4.3\,mm, 5.4\,mm, 7.0\,mm, 8.9\,mm, 11.4\,mm, 14.6\,mm, 17.7\,mm, 23.7\,mm, 30.3\,mm, 38.7\,mm, 49.3\,mm, 62.9\,mm, 80.3\,mm\} respectively.}
\label{dispersion}
\end{center}
\end{figure}

 We address the question of the impact of large scale inhomogeneities on pair dispersion. 
 Figure~\ref{dispersion_figure} shows the principle of the pair dispersion: we study the evolution of the separation between pairs of particles with time by defining
\begin{eqnarray}
D^2_{\Delta_0}(t) & = & \left\langle\left(\overrightarrow{\Delta}(t) - \overrightarrow{\Delta_0}\right)^2 \right\rangle_0, \label{eq:pairdispersion}\\
R^2_{\Delta_{0},k}(t) & = & \left\langle\left({\Delta_k}(t) - {\Delta_{0,k}}\right)^2 \right\rangle_0,
\end{eqnarray}
\noindent where $\overrightarrow{\Delta}(t)$ and $\overrightarrow{\Delta_0}$ are respectively the vectors connecting two particles at times $t$ and $t_0$ and $\Delta_{k}$ and $\Delta_{0,k}$ their projections along the $\overrightarrow{k}$ axis ($\overrightarrow{k} = \overrightarrow{x}, \overrightarrow{y}, \overrightarrow{z}$); $\langle\cdot\rangle_0$ represents the statistical mean over pairs with an initial separation $\Delta_0$ (see Fig.~\ref{dispersion}). 

The leading term of the Taylor development at short time of Eq.~\ref{eq:pairdispersion} leads to a ballisitic (or Batchelor \cite{batchelor1950}) regime: 
\begin{equation}
D^2_{\Delta_0} =  S^2_{\overrightarrow{v}}(\Delta_0) t^2 + \mathcal{O}(t^3)
\label{eq:ballistic}
\end{equation}
\noindent where $S^2_{\overrightarrow{v}}(\Delta_0)$ is the second-order Eulerian velocity structure function for a separation $\Delta_0$. 
This regime is expected to last for times $t\lesssim t^*=(\Delta_0^2/\langle \epsilon\rangle)^{1/3}$. 
For time larger than $t^*$ the super-diffusive (or Richardson-Obukhov \cite{richardson1926}) regime appears \cite{bourgoin2006}. 
It corresponds to a $t^3$ dependence of $D^2_{\Delta_0}(t)$ and was observed in turbulent convection both numerically \cite{schumacher2008} and experimentally \cite{ni2013}. 

At a given $\langle\epsilon\rangle$, the observation of the super-diffusive regime requires several conditions. 
(i) $t^*$ must be significantly smaller than the Lagrangian correlation time $T_L$, otherwise the ballistic regime transitions directly to the diffusive regime. 
(ii) Experimental tracks of particle pairs must be longer than $t^*$. 
Both conditions imply that the observation of the Richardson-Obukhov regime is most favourable for small initial separations $\Delta_0$, as confirmed in numerical simulations for both isothermal \cite{bitane2012} and thermal \cite{schumacher2008} turbulence.  
Experimentally, in high Reynolds turbulence, this regime is hard to observe \cite{bourgoin2015}. 
This is mostly due to the difficulty to access long tracks and to have good statistical convergence for small initial separations (which require high seeding densities making the tracking more complex and noisy). 
In thermal convection, Ni \& Xia \cite{ni2013} used very small initial separations (0.9--1.3\,$\eta$) in a small measurement volume and observed a fleeting super-diffusive regime. 
In our case we choose to track long trajectories over a large measurement volume (up to 30\,s corresponding to 85\,$\tau_\eta$) but with initial separations starting from 2.7\,$\eta$ ($\eta\approx0.7$\,mm).

We present in Fig.~\ref{dispersion}~(a) and (b) the mean-square separation $D^2_{\Delta_0}$ versus $t/t^*$ for short (range 1.9--2.9\,mm) and larger (range 4.3\,mm--80.3\,mm) initial separations respectively. 
In terms of Kolmogorov scale, the ranges are respectively [$2.7\,\eta$--$4.0\,\eta$] and [$6.0\,\eta$--$114\,\eta$].
For all initial separations we observe the early ballistic ($t^2$) regime. 
For the shortest separations (Fig.~\ref{dispersion}\,(a)), the trajectories are long enough compared to $t^*$ to observe the transition towards the super-diffusive ($t^3$) regime. 
This regime will be discussed further in this paper. 
For the larger initial separations (Fig.~\ref{dispersion}\,(b)), the Richardson-Obukhov regime could not be reached. 

\begin{figure}[htp]
\begin{center}
\includegraphics[width=7cm]{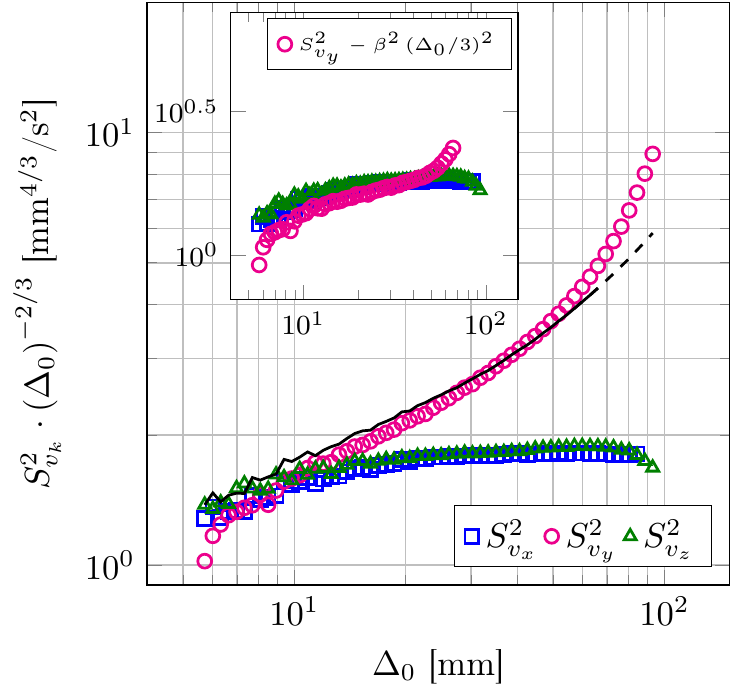}
\caption{Second-order Eulerian structure functions of each velocity component versus the total initial separation $\Delta_{0}$. 
The black line represents the model from eq.~\ref{eq:fit_dispersion}. 
The dashed part corresponds to the range where the model validity decreases. 
{In the insert, the same axes and data are plotted, except for $S^2_{v_y}$ for which the LSC term $\beta^2\Delta_0^2/3$ from eq.~\ref{eq:fit_dispersion} is subtracted to isolate the small-scale turbulent contribution. 
Only the part corresponding to the solid black line in the main plot is shown in the insert.
}}
\label{fns_struct}
\end{center}
\end{figure}

As described by Eq~\ref{eq:ballistic}, the short term pair dispersion is dominated by the $t^2$-(ballistic) regime. 
The pre-factor is given by the second-order Eulerian structure function $S_{\overrightarrow{v}}$ which in non-homogeneous thermal flows, embeds both the statistics of the turbulent spatial fluctuations and the spatial inhomogeneities of the LSC. 
We compute the structure functions with a $\Delta_{0}$ separation for each velocity component from: 
\begin{equation}
R^2_{\Delta_{0,k}}(t) = S^2_{v_k}(\Delta_{0}) t^2 + \mathcal{O}(t^3),
\end{equation}
\noindent where $S^2_{v_k}(\Delta_{0})$ corresponds to the Eulerian second-order longitudinal structure function of the $k$-component of the velocity for an initial separation $\Delta_{0}$. 

Figure~\ref{fns_struct} shows these structure functions compensated by $\Delta_{0}^{2/3}$ following classic HIT scaling. 
We observe a strong anisotropy. 
$S^2_{v_x}(\Delta_{0})$ and $S^2_{v_z}(\Delta_{0})$ present the same plateau which is in agreement with the Kolmogorov scaling, $S^2_{v_k}~\propto~\Delta_{0}^{2/3}$ \cite{monin2007}. 
From this plateau ($\SI{1.8\pm 0.1}{mm^{4/3}/s^2}$) we can estimate the local average kinetic energy dissipation rate $\langle\epsilon\rangle_\mathrm{loc}$ in the sub-domain delimited by the measurement volume -- instead of $\langle \epsilon\rangle$ for the whole cell computed using Eq~\ref{eq:epsilon}. 
Indeed, in the Kolmogorov theory we have $S^2_{v_k}=(11/9)C_2\langle\epsilon\rangle_\mathrm{loc}^{2/3}$ with $C_2$ the Kolmogorov constant \cite{monin2007}. 
In our case of weak turbulence -- the Reynolds number based on the Taylor microscale reaches $\mathrm{R}_\lambda\approx 75$, we take $C_2=1.7$ \cite{sreenivasan1995}. 
We obtain $\langle\epsilon\rangle_\mathrm{loc}=0.8\pm0.05\times10^{-6}$\,m$^2$/s$^3$. 
We can compare this value to the estimation of $\langle\epsilon\rangle$ in the whole cell using Eq~\ref{eq:epsilon}, corrected by inhomogeneity effects. 
Kunnen \emph{et al.} \cite{kunnen2008} performed numerical simulations in a cylindrical cell for $\mathrm{Pr}=6.4$ and $\mathrm{Ra}=1\times10^9$. 
From this work we estimate that in the center of the cell $\langle\epsilon\rangle_\mathrm{loc}$ is about 20\,\% to 30\,\% of its mean value calculated with Eq.~\ref{eq:epsilon}.
Assuming these corrections we {choose $\langle\epsilon\rangle_\mathrm{loc} = (0.25\pm0.05)\,\langle\epsilon\rangle$ and we finally} have $\langle\epsilon\rangle_{loc} = 1.3\pm0.3\times10^{-6}$\,m$^2$/s$^3$, which is quite consistent with the experimental deductions. 
Note that the kinetic energy dissipation rate value used to estimate $t^*$ is computed as 25\% of the value given by Eq~\ref{eq:epsilon} -- $\langle\epsilon\rangle_\mathrm{loc}\approx 1.3\times10^{-6}\,$m$^2$/s$^3$.

\begin{figure}[htp]
\begin{center}
\includegraphics[width=7cm]{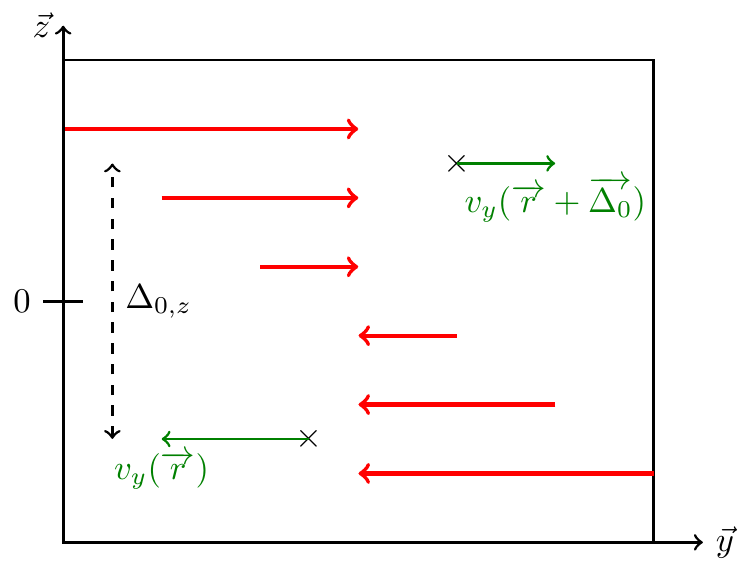}
\caption{Sketch of the mean flow in the \{$\vec{y},\vec{z}$\} plane. 
The red arrow represents the mean streamlines. 
The green arrows represent the velocity vectors of two particles separated by $\Delta_{0}$.}
\label{model_meanflow}
\end{center}
\end{figure}

$S^2_{v_y}(\Delta_{0})$ is dramatically different. 
It does not match with the Kolmogorov theory at all, but can be explained by the influence of the mean flow. 
First we decompose the velocity components{, in a manner similar to previous Lagrangian works in von K\' arm\'an and thermal turbulent flows \cite{machicoane2016,liot2017}. 
The flow velocity at a given time $t$ and position $(x,y,z)$ can be seen as the superposition of the time average velocity at this position and a time-dependent fluctuation. 
Since we use tracers, their velocity at a given time and position can be decomposed in the same way. 
Practically, the local time average velocity corresponds to the pseudo-Eulerian velocity $v_k^E$ whose computation is explained at the end of section \ref{section:setup}. 
The Lagrangian fluctuation $v_k'(t)$ is the difference between actual particle velocity and pseudo-Eulerian velocity at the particle position. 
This can be formulated as:}
\begin{equation}
v_k(t)=v^E_k(x(t),y(t),z(t))+v'_k(t),
\label{eq:decompo}
\end{equation}
\noindent where $k=x,y,z$. Using this decomposition, we can develop the structure function $S^2_{v_y}$ as:
\begin{widetext}
\begin{eqnarray}
S_{v_y}^2(\Delta_{0}) & = & \Biggl\langle\left(v_y^E(\overrightarrow{r}+\overrightarrow{\Delta_{0}})-v_y^E(\overrightarrow{r})\right)^2\Biggr\rangle_{\overrightarrow{r}}+\Biggl\langle\left(v'_y(\overrightarrow{r}+\overrightarrow{\Delta_{0}})-v'_y(\overrightarrow{r})\right)^2\Biggr\rangle_{\overrightarrow{r}}\label{eq:struct_fn} \\
 & & +2\,\Biggl\langle\left(v_y^E(\overrightarrow{r}+\overrightarrow{\Delta_{0}}-v_y^E(\overrightarrow{r})\right)\left(v'_y(\overrightarrow{r}+\overrightarrow{\Delta_{0}})-v'_y(\overrightarrow{r})\right)\Biggr\rangle_{\overrightarrow{r}},\nonumber
\end{eqnarray}
\end{widetext}
\noindent where $\overrightarrow{r}$ is a position in space. 
{Since $S^2_{v_y}(\Delta_0)$ is an Eulerian quantity, $\langle\cdot\rangle_{\overrightarrow{r}}$ represents the spatial average over all the accessible positions.} 
The third term {of Eq.~\ref{eq:struct_fn}} is a cross-correlation term between the local mean flow and the fluctuations. 
In a previous study \cite{liot2017}, a sensor-embedded particle was used to explore the flow in a parallelepipedic cell for similar $\mathrm{Pr}$ and $\mathrm{Ra}$. 
We observed that the correlations between the mean flow and the Lagrangian fluctuations are very small compared to the auto-correlations of the mean flow and the fluctuations. 
Consequently we neglect the third term. 
The second term corresponds to the second order Eulerian structure function of the fluctuations. 
We have seen (section II), for $k=x, z$ that $v_{k}^E\approx 0$, which means that $v_{k}\approx v'_{k}$ according to Eq~\ref{eq:decompo}.
Consequently, $S^2_{v_{k}}\approx\Biggl\langle\left(v'_{k}(\overrightarrow{r}+\overrightarrow{\Delta_{0}})-v'_{k}(\overrightarrow{r})\right)^2\Biggr\rangle_{\overrightarrow{r}}$. 
We also assume isotropy of turbulent fluctuations as supported by our recent investigation of single particle statistics \cite{liot2017}. 
The large similarity of $S^2_{v_x}$ and $S^2_{v_z}$ in Fig~\ref{fns_struct} reinforces the hypothesis for two-particle statistics. 
From this observation, we assume that the second term of $S^2_{v_y}$ from Eq.~\ref{eq:struct_fn} is similar to $S^2_{v_x}$:
\begin{widetext}
\begin{equation}
\Biggl\langle\left(v'_y(\overrightarrow{r}+\overrightarrow{\Delta_{0}})-v'_y(\overrightarrow{r})\right)^2\Biggr\rangle_{\overrightarrow{r}}\approx \Biggl\langle\left(v'_x(\overrightarrow{r}+\overrightarrow{\Delta_{0}})-v'_x(\overrightarrow{r})\right)^2\Biggr\rangle_{\overrightarrow{r}}\approx S^2_{v_x}.
\end{equation}
\end{widetext}
 Finally, the first term of Eq.~\ref{eq:struct_fn} is related to the mean flow structure. 
 As we observe in Fig~\ref{vy_3D}, inside our measurement volume the mean flow is mostly a shear flow in the $\{\vec{y},\vec{z}\}$ planes for every $x$, with zero-velocity in the center of the cell. 
 Thus we have $v_y^E(\overrightarrow{r})=\beta z$ (within a constant) where $\beta$ is a shear rate {defined from the $y$-component velocity gradient in the $z$ direction (which is about uniform). 
 W}e estimate it to $\beta\approx 0.14$\,s$^{-1}$ from mean vertical velocity profile. 
 The mean flow structure is sketched in Fig.~\ref{model_meanflow}. 
 Then the first term in Eq.~\ref{eq:struct_fn} can be written as: 
\begin{equation}
\Biggl\langle\left(v_y^E(\overrightarrow{r}+\overrightarrow{\Delta_{0}}))-v_y^E(r)\right)^2\Biggr\rangle_{\overrightarrow{r}} = \beta^2 \Delta_{0,z}^2.
\end{equation}
\noindent Furthermore, the initial pair separations are assumed to have three similar $\Delta_{0,k}$ in order to avoid considering particles with a very large separation in one direction and a very short separation in an other one. 
Consequently we have $\Delta_0\approx\sqrt{3}\Delta_{0,z}$. 
Finally we can write: 
\begin{equation}
S_{v_y}^2(\Delta_{0})\approx\beta^2\left(\frac{\Delta_{0}^2}{3}\right) + S_{v_x}^2(\Delta_{0}).
\label{eq:fit_dispersion}
\end{equation}

The black line in Fig.~\ref{fns_struct} compares this no-free-parameter model to experimental data. 
It is valid while the three components of the initial separation are similar. 
Since the volume is larger in the $\vec{z}$ direction than in the horizontal ones, this condition is violated for $\Delta_0 \gtrsim 50$\,mm. 
The solid part of the line corresponds to the zone where the model is expected to be valid, whereas the dashed part, which departs from experimental data, corresponds to a range of initial separations where the condition $\Delta_0\approx\sqrt{3}\Delta_{0,k}$ is not satisfied anymore. 
{Fig.~\ref{fns_struct} and its insert show the same experimental results, except for $S^2_{v_y}$.
In the insert, the mean-flow component in Eq.~\ref{eq:fit_dispersion}, $\beta^2\Delta_0^2/3$ is subtracted from  experimental data $S^2_{v_y}$. 
This leads to a collapse of the curve on the $S^2_{v_x}$ and $S^2_{v_z}$ on the model validity domain. 
This collapse shows that the Eulerian second-order structure function $S^2_{v_y}$ recovers the HIT behaviour if the mean flow influence is removed.}

\section{About the Richardson-Obukhov regime}

\begin{figure}[htp]
\begin{center}
\includegraphics[width=7.5cm]{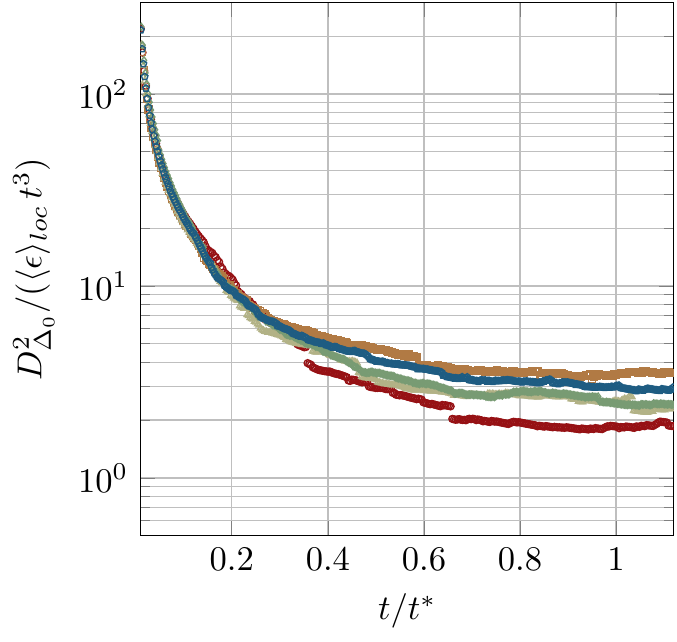}
\caption{Compensated plot of the pair dispersion $D^2_{\Delta_0}/(\langle \epsilon \rangle_\mathrm{loc} t^3)$ for short initial separations: 2.7\,$\eta$, 3.0\,$\eta$, 3.3\,$\eta$, 3.7\,$\eta$, 4.0\,$\eta$.}
\label{richardson}
\end{center}
\end{figure}

{The observation of the Richardson-Obukhov regime is experimentally subtle in turbulent convection. 
Moreover, the super-diffusive regime could include extra diffusion due to the LSC-generated shear rate.} 
For short initial separations, we observe in Fig.~\ref{dispersion}\,(a) a transition towards a $t^3$ regime similar to that of Richardson-Obukhov. 
In this regime, the expected pair separation expression is: 
\begin{equation}
D^2_{\Delta_0}=g\langle\epsilon\rangle_\mathrm{loc} t^3,
\end{equation}
\noindent where $g$ is called the Richardson constant \cite{salazar2009,bourgoin2015}. 
In homogeneous isotropic turbulence (HIT), the expected value is $g\in[0.65-0.7]$ in the range of Reynolds number $R_\lambda$ corresponding to our experiment \cite{salazar2009}. 
Figure~\ref{richardson} shows the pair separation compensated by $\langle\epsilon\rangle_\mathrm{loc}\,t^3$ for the smallest initial separations (corresponding to Fig.~\ref{dispersion}\,(a)). 
A plateau is well defined for $t$ between 0.7\,$t^*$ and 1.1\,$t^*$ with $g\in [2-4]$. 
This range is consistent although a bit larger than HIT value. 
However it is significantly larger than values reported in turbulent convection by Ni \& Xia \cite{ni2013} ($g\approx 0.1$ for $\Delta_0\in [0.9\,\eta-1.3\,\eta]$). 
Moreover, the plateau is observed for lower $t/t^*$ than in Ni \& Xia. 
Aditionally we find a systematic dependency on the plateau in Fig.~\ref{richardson} with the initial separation, $\Delta_0$.
 
\begin{figure}[htp]
\begin{center}
\includegraphics[width=7.5cm]{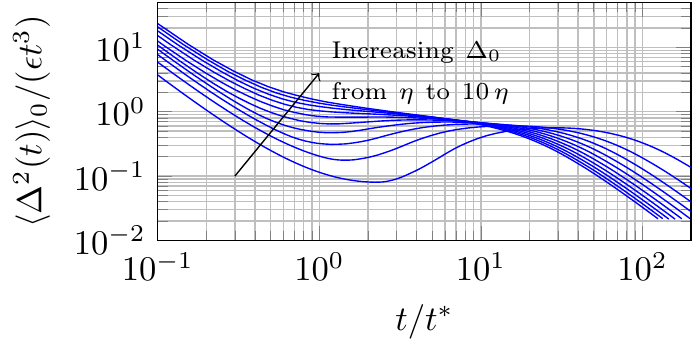}
\caption{Pair separation $\langle\Delta^2 (t)\rangle_0$ for HIT computed from the ballistic-cascade model (adapted to account for the dissipative scaling of $S_2$ at small initial separations) \cite{bourgoin2015}, compensated by $(\epsilon t^3)$. 
The initial separations rise linearly in the range $[\eta,10\,\eta]$.}
\label{separation_numeric}
\end{center}
\end{figure}

 Both observations can be explained from the peculiar behaviour of pair dispersion for initial separation close to the dissipative scale, and by the value of the average kinetic energy dissipation rate used. 
 The shift in the time value for the occurrence of the plateau between our data and that of Ni \& Xia can be explained by our use of $\langle\epsilon\rangle_\mathrm{loc}$ to compute $t^*$ which leads to a higher $t^*$ than when using $\langle\epsilon\rangle$. 
 Concerning the plateau value, numerical simulations performed by Boffetta \& Sokolov \cite{boffetta2002} and Sawford \emph{et al.} \cite{sawford2001,sawford2008} show that for initial separations of the order $\eta$, the $t^3$ {Richardson-Obukhov} regime is preceded by a local minimum, leading to {an apparent} short $t^3$ lower plateau. 
 This phenomenon disappears for higher $\Delta_0$. 
 This is attributed to a contamination of the initial range by dissipation effects \cite{sawford2001,boffetta2002}. 
 Based on the ballistic cascade model proposed by Bourgoin \cite{bourgoin2015}, we plot in Fig.~\ref{separation_numeric} the pair separation compensated by $(\epsilon t^3)$ in HIT for initial separations between $\eta$ and 10\,$\eta$. 
 The local minimum zone before the super-diffusive plateau is clearly visible for an initial separation $\Delta_0\sim\eta$. 
 A progressive disappearance of the minimum zone is observed as $\Delta_0$ increases. 
 This highlights that tracks too short can lead to a biased {(misleadingly too small)} estimation of the Richardson constant, misleadingly taken as the apparent plateau observed for small $\Delta_0$ near the local minimum between ballistic and super-diffusive zone in the $D^2_{\Delta_0}/(\langle\epsilon\rangle_\mathrm{loc}t^3)$ plot. 
 In this zone the apparent plateau also leads to an erroneous dependency of $g$ on $\Delta_0$. 
 Figure~\ref{separation_numeric} suggests that the plateau due to the local minimum zone is nearly level with the actual Richardson plateau for $\Delta_0\gtrsim 4\,\eta$. 
 These observations very likely explain the apparent initial separation dependency observed in Fig.~\ref{richardson} and {also} reported in previous work by Ni \& Xia \cite{ni2013}. 
 As in both studies, due to the limited track length, it is likely that the local minimum zone is explored, rather than the actual plateau of the Richardson regime. 
 Ni \& Xia proposed an estimation of the Richardson constant based on the smallest initial separation they had ($\Delta_0\lesssim\eta$), for which the plateau related to the local minimum zone significantly under-estimates the actual value of $g$. 
 In this previous pioneering study, the tracks for an initial separation larger than $\eta$ {were too short} and not usable to estimate $g$.
 In our study, the tracks {of pairs with initial separations} in the range $[2.7\,\eta-4\,\eta]$ are marginally longer and allow for computation of the plateau due to the local minimum zone.
 Since we have larger initial separations, the plateau related to the local minimum zone is {naturally} higher (see Fig.~\ref{separation_numeric}) than for Ni \& Xia. 
 {Based on the previous discussion, the plateau around the minimum zone roughly levels to the actual Richardson plateau for initial separations close to $4\eta$.
 Considering the range of initial separations in Fig.~\ref{richardson} ($[2.7\,\eta-4\,\eta]$) we can therefore expect that the observed plateau leads to a reasonable estimate of the actual Richardson constant, hence $g \in [2-4]$.}
 Note that the use of $\langle\epsilon\rangle$ instead of $\langle\epsilon\rangle_\mathrm{loc}$ also further under-estimates the plateau value in the work by Ni \& Xia \cite{ni2013}.

 \section{Discussion and conclusion}  
 
 {To explore the influence of inhomogeneity and ani\-sotropy on turbulent statistics, a}n experimental study of pair dispersion in a turbulent thermal flow was performed. 
 Our experimental setup and analysis/post-processing tools allow us to obtain long trajectories, compared to the Kolmogorov time, and exceeding Batchelor time $t^*$ in some cases.
 
 The quantitative analysis of the ballistic regime of pair dispersion highlights the influence of the Large Scale Circulation on the turbulent transport and gives a way to measure the mean kinetic energy dissipation rate in the considered volume. 
 {Given its spatial inhomogeneity, a measurement of this rate is necessary to compare thermal convection turbulence to HIT.} 
 This assessment is in good agreement with an estimation from global and Eulerian approaches. 
 We also used pair dispersion to access Eulerian velocity structure functions using only particle displacement (without needing to derive their velocity). 
 {This is a way to study the influence of the inhomogeneity on thermal flow turbulence.} 
 The particle dispersion in the $\vec{y}$ direction is highly influenced by the convection roll. 
 This is visible on the Eulerian velocity structure function which departs from the $\Delta_0^{2/3}$ Kolmogorov scaling. 
 We proposed a model to describe the shape of this structure function which mixes a phenomenological approach of the mean flow, and experimental data from the other horizontal velocity structure functions, not affected by the mean flow. 
 This last point is useful to take into account the shape of the velocity structure function at low initial separations before reaching the Kolmogorov scaling. 
 With this choice, the non-negligible viscous dissipation for initial separations close to the Kolmogorov scale is considered. 
 This model is in good agreement with experimental data, except for large initial separations, due to the loss of the hypothesis of equipartition for initial separation components. 
 {While turbulent convection is intrinsically inhomogeneous and anisotropic, this approach demonstrates that we can recover statistics from usual HIT simply by removing the mean flow.
 This is not trivial, especially for two-particle statistics. 
 This also shows that there is no temperature influence on the turbulence organisation, which is important for understanding the temperature role in thermal turbulent flows. 
 Our experiments validate this point in the center part of the convection cell, where plumes are scarce. }
 
 For the smallest initial separations, we obtain trajectories longer than $t^*$ and we are able to observe the transition from ballistic to super-diffusive regime with a Richardson constant comparable but larger than HIT. 
 {For short initial separations, this non-trivial transition reveals some complex behaviours even in HIT. 
 Moreover, the extra diffusion due to LSC-generated shear rate could affect the observations of the super-diffusive regime.} 
 The difference with other experimental results could be explained from a well-predicted short plateau due to a local minimum zone which appears between ballistic and super-diffusive zones at short initial separations on compensated plots.
 
{To summarise, our study shines a light on three new physical insights about turbulent thermal convection.  Using a decomposition between Large Scale Circulation and turbulent fluctuations, we can compute the relative impact of each contribution to pair dispersion. Then, we compute Eulerian second-order velocity structure functions from pair separations. Using the same decomposition to remove LSC contribution, we reveal that the remaining statistics recover usual HIT behaviours, being careful to use a local estimation of the kinetic energy dissipation rate. Finally, we propose a revisited and more precise estimation of the Richardson constant. }
 
In addition to {these points}, the pair separation is a good statistical tool to study the transport properties of turbulent thermal flows, especially in the presence of an inhomogeneous mean flow. 
Because of the large field of view in our experiment, we are able to study the effects of inhomogeneity due to Large Scale Circulation specific to the Rayleigh-B\'enard convection. 
In future studies we aim to explore an even wider measurement volume and be able to study the jets where a high concentration of thermal plumes is observed. 
These coherent structures should affect the different components of the pair dispersion \cite{schumacher2008}. 
Furthermore, measurements of particle dispersion in the jets could give information on thermal transfer in the Rayleigh-B\'enard convection. 
Possible analysis would look at the next order of the Taylor expression of Eq.~\ref{eq:ballistic}. 
This can be done by computing the difference between the forward-in-time and the backward-in-time pair dispersions \cite{jucha2014}, and would open new perspectives to study the energy cascade in turbulent convection.

\begin{acknowledgments}
  We gratefully thank Marc Moulin for manufacturing of the cell and Miguel L\'opez-Caballero for his help with the camera calibration. 
  The data processing and analysis were made possible with the PSMN computer resources. 
  We warmly acknowledge Theresa Oehmke for the manuscript proofreading.
\end{acknowledgments}
 
\bibliography{biblio_tolosa_2particules}

\end{document}